\documentclass[conference]{IEEEtran}
\usepackage[
  pass,
]{geometry}

\usepackage{cite}
\usepackage[numbers]{natbib} 

\ifCLASSINFOpdf
\else
\fi
\usepackage[usenames,dvipsnames,svgnames,table]{xcolor} 
\PassOptionsToPackage{hyphens}{url}
\usepackage{hyperref}
\usepackage{multirow}
\usepackage[export]{adjustbox} 
\usepackage{tabularx}
\usepackage{booktabs}
\usepackage[framemethod=tikz]{mdframed}
\newmdenv[
  innerleftmargin=7pt,
  innerrightmargin=7pt,
  tikzsetting={draw=black,dashed,line width=0.5pt,dash pattern = on 4pt off 2pt},
  linecolor=white,
  backgroundcolor=white
]{dashedbox}
\newmdenv[
  innerleftmargin=7pt,
  innerrightmargin=7pt,
  tikzsetting={draw=black, line width=0.5pt},
  linecolor=black,
  backgroundcolor=white
]{normalbox}
\let\labelindent\relax
\usepackage{enumitem}

\makeatletter
\let\MYcaption\@makecaption
\makeatother
\usepackage[font=footnotesize]{subcaption}
\makeatletter
\let\@makecaption\MYcaption
\makeatother

\usepackage[export]{adjustbox} 
\usepackage{amssymb,amsmath}

\newcommand{\mycode}[1]{{\small\texttt{#1}}}

\usepackage{array}
\newcolumntype{L}[1]{>{\raggedright\let\newline\\\arraybackslash\hspace{0pt}}m{#1}}
\newcolumntype{C}[1]{>{\centering\let\newline\\\arraybackslash\hspace{0pt}}m{#1}}
\newcolumntype{R}[1]{>{\raggedleft\let\newline\\\arraybackslash\hspace{0pt}}m{#1}}

\begin{document}
%
\title{Navigate, Understand, Communicate:\\ How Developers Locate Performance Bugs}



%
\author{
\IEEEauthorblockN{
Sebastian Baltes\IEEEauthorrefmark{1},
Oliver Moseler\IEEEauthorrefmark{1},
Fabian Beck\IEEEauthorrefmark{2}, and
Stephan Diehl\IEEEauthorrefmark{1}}
\IEEEauthorblockA{\IEEEauthorrefmark{1}
University of Trier, Germany\\
}
\IEEEauthorblockA{\IEEEauthorrefmark{2}
VISUS, University of Stuttgart, Germany\\
}}


\maketitle

\begin{abstract}
\emph{Background:} Performance bugs can lead to severe issues regarding computation efficiency, power consumption, and user experience. Locating these bugs is a difficult task because developers have to judge for every costly operation whether runtime is consumed necessarily or unnecessarily.
\emph{Objective:} We wanted to investigate how developers, when locating performance bugs, navigate through the code, understand the program, and communicate the detected issues. 
\emph{Method:} We performed a qualitative user study observing twelve developers trying to fix documented performance bugs in two open source projects. The developers worked with a profiling and analysis tool that visually depicts runtime information in a list representation and embedded into the source code view. 
\emph{Results:} We identified typical navigation strategies developers used for pinpointing the bug, for instance, following method calls based on runtime consumption. The integration of visualization and code helped developers to understand the bug. Sketches visualizing data structures and algorithms turned out to be valuable for externalizing and communicating the comprehension process for complex bugs.
\emph{Conclusion:} Fixing a performance bug is a code comprehension and navigation problem. Flexible navigation features based on executed methods and a close integration of source code and performance information support the process. 

\end{abstract}


%
\IEEEpeerreviewmaketitle

\section{Introduction}

Performance is a non-functional requirement that every software needs to fulfill, at least to some extent. It might be hard to optimize the performance of already well-implemented algorithms.
But often, bugs in form of unnecessarily complex or slow operations affect the performance of software.
 \citeauthor{Jin12}~\cite{Jin12} define \emph{performance bugs} as ``defects where relatively simple source code changes can significantly speed up software, while preserving functionality.''
In that sense, performance bugs significantly differ from usual bugs, which are deviations of the program behavior from specified functional requirements. Nonetheless, performance bugs are critical as well because they could corrupt user experience, reduce system throughput, increase latency, and waste computational resources~\cite{Jin12, Nistor13}. Although they can be fixed with simple changes, locating and understanding them is a difficult task: complex chains of executed methods need to be traced and, for every statement and branch, the developer needs to clarify whether runtime was consumed not more than appropriate.

To build tools that support developers in locating and fixing performance bugs, it is essential to understand how the debugging process works. Different studies have already been conducted that investigate fixing functional bugs~\cite{Ko06, Johnson13, Lawrance13, Beck15}. We are, however, not aware of any study that focuses on observing developers locating performance bugs. Results from other studies on debugging processes cannot be transferred directly because the steps and tools required to optimize a non-functional requirement like performance are substantially different from those applied for fixing a functional bug.
These differences include: (i) developers cannot analyze whether a program is correct regarding performance because there only exist better or worse solutions; (ii) developers need to investigate not only program state but also runtime consumption; and (iii) collecting runtime information requires to set up realistic benchmarks that differ from usual regression tests. Also, \citeauthor{Jin12}~\cite{Jin12} already pointed at the lack of studies on how performance bugs are fixed by developers.

The user study presented in this paper aims at filling this gap by investigating how developers \emph{navigate} through code, \emph{understand} performance problems, and \emph{communicate} with each other to fix performance bugs.
It is based on a visual performance analysis tool~\cite{Beck13b} that we extended to provide developers versatile means of navigation and support for comprehension (Section~\ref{sec:tool}).
The specific research questions of this paper cover activities that enable developers to fix a performance bug (Section~\ref{sec:research_questions}).
The evaluation we designed to answer these questions is an extensive qualitative user study where twelve developers fixed real-world performance bugs in pairs (Section~\ref{sec:design}).
The results are based on a detailed evaluation of interviews, recordings, and interaction logs.
They provide insights into how developers navigate to locate the bugs and communicate the detected issues to other developers (Section~\ref{sec:results}).
While the controlled setting and the detailed analysis support the credibility of these results, the relatively small number of participants and our focus on collection libraries limits their generalizability (Section~\ref{sec:validity}).
Finally, we discuss our findings in the context of previous studies on related subjects (Section~\ref{sec:related_work}) and name implications for tools and future work (Section~\ref{sec:conclusion}).

\section{Visual Performance Analysis}
\label{sec:tool}

Profiling tools record individual program runs and assign measured performance values to code entities such as methods. While performance considerations could include runtime, memory, or latency, we focus on analyzing runtime consumption as one of the most prominent performance metrics. The standard user interface for inspecting profiling results are lists of code entities that can be sorted by  runtime consumption or organized hierarchically to follow execution sequences. When exploring performance bugs, however, we assume that the source code is a valuable asset as well: developers need to switch back and forth between performance information and code to figure out why a certain amount of runtime was consumed in a specific code entity and how code entities are related to each other. Using different tools or views to explore runtime information and code could result in considerable navigation overhead and unnecessary cognitive load.
Since we want to explore how developers navigate to locate performance bugs, we provide different options for navigation: First, our study is based on IntelliJ IDEA, a popular integrated development environment (IDE) for Java that includes versatile options to navigate through a software project. Second, we provide a sortable list of code entities showing performance information comparable to standard profiling tools (Fig.~\ref{fig:list}). Third, based on an approach introduced by Beck et al.~\cite{Beck13b}, we add word-sized visualizations to augment code entities with performance information in the code editor (Fig.~\ref{fig:insitu}). This section briefly describes the specific tool used in the study, which is an extended version of the original approach by Beck et al.~\cite{Beck13b}. It is capable of profiling Java systems and uses a sampling-based profiling technique.

\begin{figure}[tbp] \centering
\includegraphics[width=0.8\columnwidth, trim=0.0in 0.0in 0.0in 0.0in]{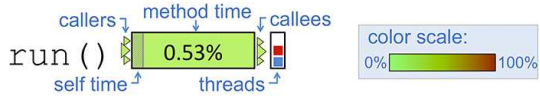}
\caption{Enlarged and annotated sparkline visualization augmenting the code of a method \texttt{run()}  with runtime information (annotations in blue).}
\label{fig:sparkline}
\end{figure}

\subsection{In-situ Visualization}

\begin{figure}[tbp] \centering
\includegraphics[width=1.0\columnwidth, trim=0.0in 0.0in 0.0in 0.0in, clip=true]{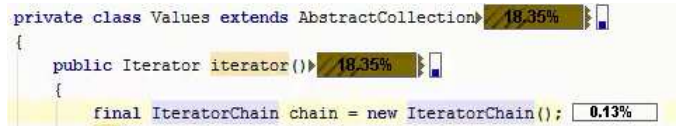}
\caption{In-situ visualization of performance information within the code view for classes, methods, and method calls.}
\label{fig:insitu}
\end{figure}

Our tool integrates word-sized visualizations, also known as \emph{sparklines}~\cite{Tufte06}, to augment the source code view of the IDE (Fig.~\ref{fig:insitu}). Other so-called \emph{in-situ software visualization} approaches~\cite{Harward10} have already been used to visualize static or dynamic software metrics~\cite{Harward10, Roethlisberger09}, to monitor numeric variables~\cite{Beck13b}, or to support feature location tasks~\cite{Beck15}. The motivation for these kinds of approaches is to avoid \emph{split-attention effects}~\cite{Ayres05,Paas05} that likely affect code navigation and comprehension whenever developers need to switch between code representation and analysis view. Providing such an in-situ approach in the evaluation will give insights whether the suggested advantages are relevant in practical application and how the approach is received by developers.

\begin{figure}[tbp] \centering
\includegraphics[width=1.0\columnwidth, trim=1.63in 0.03in 0.01in 0.0in, clip=true]{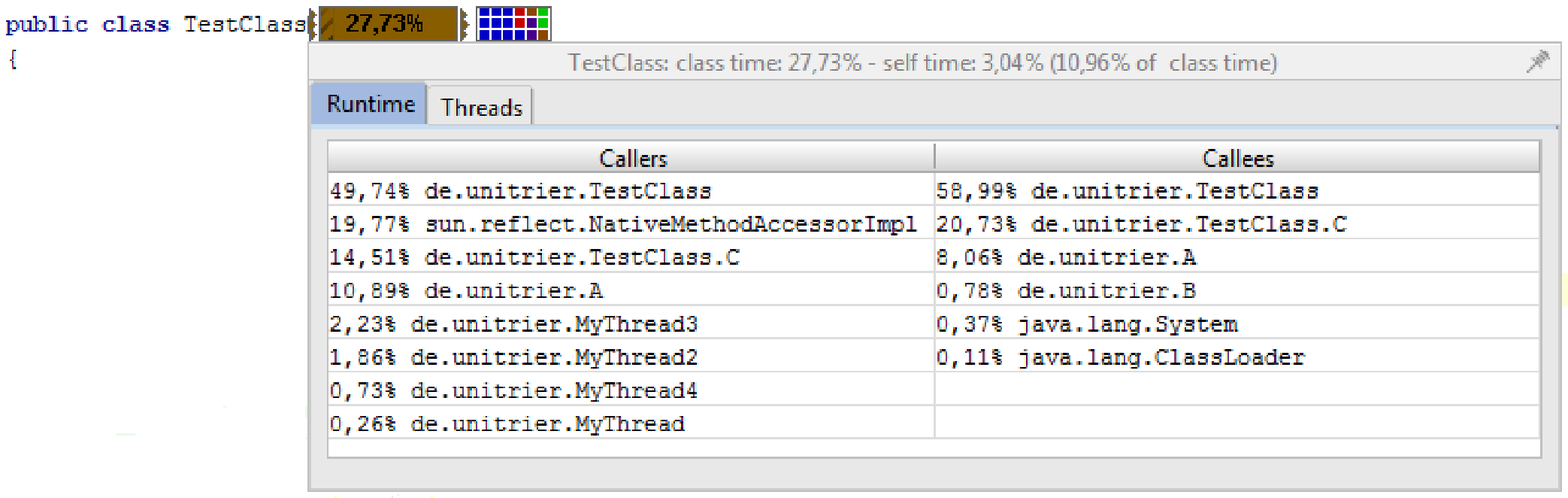}
\caption{Tooltip dialog showing details of callers and callees of a in-situ visualization for a class \texttt{TestClass}.}
\label{fig:tooltip}
\end{figure}

The in-situ approach enriches every method declaration with a word-sized sparkline visualization that summarizes the most important runtime information regarding the execution of the method in a recorded program execution (\emph{method visualization}, Fig.~\ref{fig:sparkline}).
The central information is the \emph{method time} (i.e., the percentage of total runtime the method has been active) depicted as a value and color-coded in the background of the main rectangle on a scale from light green (low) to dark red (high). As a striped part, the \emph{self time} (i.e., the part of the runtime that was actually consumed by statements of the method) is added to that rectangle. Small arrows at the left indicate how many callers (0, 1, 2, or $>$2) of the method were active; an analogous representation is added for callees on the right. Another box on the right summarizes the threads that executed the method: each little square represents an instance of a thread, the color encodes its type. More details and examples can be found in the description of the original approach~\cite{Beck13b}. 
Within the body of each method, every method call that is covered by recorded execution information is assigned a simplified in-situ visualization (\emph{method call visualization}, Fig.~\ref{fig:insitu}). It expresses the percentage of runtime that the call propagates with respect to the total \emph{method time}.
To support analysis on a higher level of abstraction, the visualizations augmenting every class declaration aggregate the runtime of all methods contained in the class (this feature has been added to the original approach~\cite{Beck13b}). Tooltip dialogs are available for all in-situ visualizations on demand and allow retrieving precise information on runtime, caller, callees, and threads (Fig.~\ref{fig:tooltip}).

\subsection{List Representation}

\begin{figure}[tbp] \centering
\includegraphics[width=1.0\columnwidth, trim=0.0in 4.49in 0.0in 0.0in, clip=true]{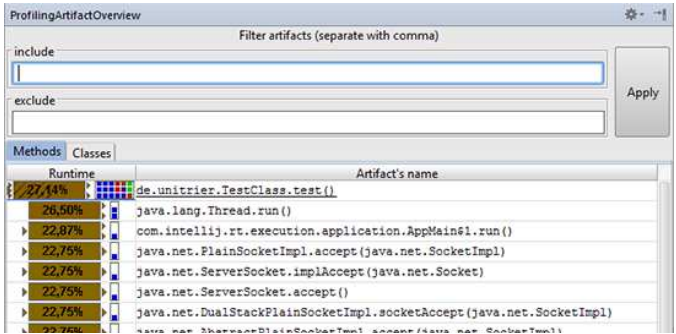}
\caption{List representation of code entities (methods and classes) sorted by consumed runtime with filter options (image cropped).}
\label{fig:list}
\end{figure}

To also provide a more traditional representation, we added a list view of performance information to the tool (Fig.~\ref{fig:list}). The list presents all executed methods or classes (selectable as tabs) sorted by total runtime consumption in decreasing order (i.e., the `hottest' code entities are listed at the top). For sake of consistency and to provide richer information, we used the same sparkline visualizations as described above to summarize the runtime information of methods and classes. Again, details of the depicted information can be retrieved as tooltip dialogs on demand (Fig.~\ref{fig:tooltip}).
To quickly filter the list, developers might define include and exclude filters by inserting terms in the corresponding text fields at the top of the view; multiple terms are connected with a logical `OR'.

\subsection{Navigation}

Navigating through the source code of the studied software system is a central task when locating performance bugs.
Therefore, we provide specialized navigation options in addition to the features already available in the IDE (e.g., project overview, class outline, method calls as hyperlinks). While the IDE navigation is limited to static call information, our extensions focus on enabling the developers to follow dynamic instances of method calls, that is, the calls actually performed during the recorded execution:

\begin{itemize}[leftmargin=*, itemsep=1ex]
\item \textbf{In-situ Visualization:} All methods listed in the tooltips of the in-situ visualizations are clickable; the respective class is opened and the editor jumps to the selected method.

\item \textbf{List Representation:} The list items are clickable and open the respective entity in the editor. Again, tooltips provide an option to jump to callers and callees of the  entity.
\end{itemize}

These features do not replace any IDE features, but extend them.
Developers are free to use (or not to use) any of the presented visualizations and navigation features for locating a performance bug.

\section{Research Questions}
\label{sec:research_questions}

From the general goal of better understanding how developers locate and fix performance bugs, we derive research questions that we address in this paper. We first formulate general questions that are too broad to be covered in a single user study. Subquestions acting as instances of the general questions focus the problem to specific issues that we answer based on results of our user study. 

When working on change tasks, like fixing performance bugs, developers do not aim at understanding a system entirely, but want to understand just enough to make the required change~\cite{Singer97}.
To this end, they navigate the source code and try to understand the relevant parts to build up their task context~\cite{Murphy05}. Information sources used for navigation and navigation strategies reflect this understanding process. We also want to learn how the additional features provided by the in-situ visualization approach influence the navigation process.

\noindent \textbf{RQ1} (Navigating and Understanding): \emph{How do developers navigate and what information and representation is supportive for locating a performance bug?}

\begin{itemize}[leftmargin=*, itemsep=1ex]
\item[] \textbf{RQ1.1} \emph{How was information from the profiling tool or other parts of the IDE used to locate the performance bug?}
\item[] \textbf{RQ1.2} \emph{Is the in-situ visualization of the profiling data 
beneficial compared to a traditional list representation?}
\item[] \textbf{RQ1.3} \emph{What navigation strategies do developers pursue to locate a specific performance bug?}
\end{itemize}

Communication can be considered as an externalization of a reasoning process. To gain deeper insights into the understanding process behind locating a performance bug, we also wanted to study how developers explain performance issues. A setting where developers communicate with each other naturally while solving a software development task is pair programming~\cite{Williams00}. As an additional externalization and medium of communication, we investigated the use of sketches for developing an understanding of performance bugs. Like pair programming, sketches are a common and natural method for collaboration in software engineering~\cite{Baltes14, Cherubini07}.

\noindent \textbf{RQ2} (Understanding and Communicating): \emph{How do developers try to understand and explain the causes of performance bugs?}

\begin{itemize}[leftmargin=*, itemsep=1ex]
\item[] \textbf{RQ2.1} \emph{How do developers communicate with each other when locating a performance bug?}
\item[] \textbf{RQ2.2} \emph{Could sketches help understand and communicate a performance bug?}
\end{itemize}

The study described in the following aims at answering these specific research questions. Conducting the study as a realistic software engineering task was a principle that guided its design.

\section{Study Design}
\label{sec:design}

We designed the study as a structured qualitative observation study in a controlled setting.
Twelve software developers participated in teams of two, resulting in six study sessions.
Each session was divided into three phases: after a tutorial, the teams were asked to locate and fix four real-life performance bugs, followed by a questionnaire collecting feedback from the participants.
After summarizing demographic data of the participants, we describe details of the study design and how the sessions were executed.

\subsection{Participants}
\label{sec:participants}

\begin{table*}[t]
	\caption{Participants (experience in object-oriented programming, Java, collections and data structures, IntelliJ, other IDEs, fixing performance bugs, our tool, and profiling tools in general)}
	\label{tab:participants}
	\centering
	\begin{tabular}{ccccC{0.85cm}C{0.85cm}C{0.85cm}C{0.85cm}C{0.85cm}C{1cm}C{1cm}C{0.85cm}}
		\hline
		Team	& Participant		& Current Occupation		& Work Exp.			& \multicolumn{8}{c}{Experience (no exp. = 0 to 4 = expert)} \\
				&				&						& (years)			& \multicolumn{1}{c}{OOP} & \multicolumn{1}{c}{Java} & \multicolumn{1}{c}{Collec.} & \multicolumn{1}{c}{IntelliJ} & \multicolumn{1}{c}{IDEs} & \multicolumn{1}{c}{Perf.Bugs} & \multicolumn{1}{c}{Our Tool}  & \multicolumn{1}{c}{Profiling} \\ 
		\hline
		\multirow{2}{*}{T1}	& P1 		& Research assistant				& 5		& 4\cellcolor{DeepSkyBlue} & 4\cellcolor{DeepSkyBlue} & 3\cellcolor{LightSkyBlue} & 3\cellcolor{LightSkyBlue} & 3\cellcolor{LightSkyBlue} & 1\cellcolor{LightSalmon} & 1\cellcolor{LightSalmon} & 0\cellcolor{RedOrange} \\ 
							& P2		& Research assistant				& 5		& 4\cellcolor{DeepSkyBlue} & 4\cellcolor{DeepSkyBlue} & 4\cellcolor{DeepSkyBlue} & 1\cellcolor{LightSalmon} & 4\cellcolor{DeepSkyBlue} & 2\cellcolor{LightGray} & 1\cellcolor{LightSalmon} & 1\cellcolor{LightSalmon} \\ 
		\hline
		\multirow{2}{*}{T2}	& P3		& MSc student, industry exp.		& 1		& 3\cellcolor{LightSkyBlue} & 3\cellcolor{LightSkyBlue} & 2\cellcolor{LightGray} & 0\cellcolor{RedOrange} & 3\cellcolor{LightSkyBlue} & 1\cellcolor{LightSalmon} 	& 0\cellcolor{RedOrange} & 2\cellcolor{LightGray} \\ 
							& P4		& MSc student, industry exp.	 	& 3		& 3\cellcolor{LightSkyBlue} & 3\cellcolor{LightSkyBlue} & 3\cellcolor{LightSkyBlue} & 1\cellcolor{LightSalmon} & 2\cellcolor{LightGray} & 1\cellcolor{LightSalmon} & 0\cellcolor{RedOrange} & 1\cellcolor{LightSalmon} \\ 
		\hline
		\multirow{2}{*}{T3}	& P5 		& Software developer		& 3		& 4\cellcolor{DeepSkyBlue} & 3\cellcolor{LightSkyBlue} & 4\cellcolor{DeepSkyBlue} & 1\cellcolor{LightSalmon} & 3\cellcolor{LightSkyBlue} & 3\cellcolor{LightSkyBlue} & \cellcolor{LightSalmon}1 & 2\cellcolor{LightGray} \\ 
							& P6		& Diploma student			& 4		& 3\cellcolor{LightSkyBlue} & 3\cellcolor{LightSkyBlue} & 3\cellcolor{LightSkyBlue} & 4\cellcolor{DeepSkyBlue} & 2\cellcolor{LightGray} & 1\cellcolor{LightSalmon} & 1\cellcolor{LightSalmon} & 0\cellcolor{RedOrange} \\ 
		\hline
		\multirow{2}{*}{T4}	& P7		& MSc student			& 0			& 3\cellcolor{LightSkyBlue} & 2\cellcolor{LightGray} & 3\cellcolor{LightSkyBlue} & 1\cellcolor{LightSalmon} & 2\cellcolor{LightGray} & 1\cellcolor{LightSalmon} & 0\cellcolor{RedOrange} & 0\cellcolor{RedOrange} \\ 
							& P8		& MSc student			& 0			& 1\cellcolor{LightSalmon} & 1\cellcolor{LightSalmon} & 0\cellcolor{RedOrange} & 0\cellcolor{RedOrange} & 1\cellcolor{LightSalmon} & 1\cellcolor{LightSalmon} & 0\cellcolor{RedOrange} & 1\cellcolor{LightSalmon} \\ 
		\hline
		\multirow{2}{*}{T5}	& P9		& Research assistant, industry exp.	& 10		& 3\cellcolor{LightSkyBlue} & 2\cellcolor{LightGray} & 3\cellcolor{LightSkyBlue} & 0\cellcolor{RedOrange} & 4\cellcolor{DeepSkyBlue} & 4\cellcolor{DeepSkyBlue} 	& 0\cellcolor{RedOrange} & 3\cellcolor{LightSkyBlue} \\ 
							& P10	& Research assistant, industry exp.	& 6			& 2\cellcolor{LightGray} & 2\cellcolor{LightGray} & 2\cellcolor{LightGray} & 3\cellcolor{LightSkyBlue} & 1\cellcolor{LightSalmon} & 3\cellcolor{LightSkyBlue} & 0\cellcolor{RedOrange}		& 2\cellcolor{LightGray} \\ 
		\hline
		\multirow{2}{*}{T6}	& P11 	& Software developer	& 15		& 3\cellcolor{LightSkyBlue} & 1\cellcolor{LightSalmon} & 3\cellcolor{LightSkyBlue} & 0\cellcolor{RedOrange} & 3\cellcolor{LightSkyBlue} & 2\cellcolor{LightGray} & 1\cellcolor{LightSalmon} & 2\cellcolor{LightGray} \\ 
							& P12	& Software developer	& 1			& 3\cellcolor{LightSkyBlue} & 3\cellcolor{LightSkyBlue} & 2\cellcolor{LightGray} & 2\cellcolor{LightGray} & 2\cellcolor{LightGray} & 1\cellcolor{LightSalmon} & 0\cellcolor{RedOrange} & 1\cellcolor{LightSalmon} \\ 
		\hline
							&		& mean values:	& 4.4		& 3.0\cellcolor{LightSkyBlue} & 2.6\cellcolor{LightSkyBlue} & 2.7\cellcolor{LightSkyBlue} & 1.3\cellcolor{LightSalmon} & 2.5\cellcolor{LightSkyBlue} & 1.8\cellcolor{LightGray} & 0.4 \cellcolor{RedOrange} & 1.3\cellcolor{LightSalmon} \\ 
		\hline
	\end{tabular}

\end{table*}

The twelve software developers who participated in our study were all male and between 22 and 43 years old (mean value: 30.7 years). 
As summarized in Table~\ref{tab:participants}, three of them worked as software developers, four were research assistants at a computer science department (two of them worked full-time in the software industry before), and five were graduate students enrolled in a computer science program (two of them were working half-time as software developers in industry). In total, ten participants had professional work experience in software development (between 1 and 15 years).
We asked the participants to rate their experience in different areas related to the study on Likert items ranging from 0 (no experience) to 4 (expert). 
Summarizing the ratings, participants had a good level of experience in object-oriented programming, the Java programming language, collections and data structures, and working with integrated development environments.
Four participants had no experience with the IntelliJ IDEA IDE.
However, this did not seem to cause any problems during the study, presumably because modern IDEs have quite similar features and our participants got used to the unfamiliar IDE quickly. 
Every participant had at least some experience in fixing performance bugs, but only five participants were experienced (experience $\ge 2$).
Most of them did not know our profiling and visualization tool before. The experience with profiling tools in general was rather low.
Nevertheless, in each team, there was at least one member having some experience with profiling tools.

\subsection{Procedure}

The study procedure was divided into three phases. During the whole study, at least one of the authors was present to help participants in case of questions. All tasks were executed in the same order by all teams.

\noindent\textbf{1) Tutorial:} In the first phase, each pair was introduced to profiling in general, the sampling and analysis approach that our tool uses, and various features of the tool.
To this end, we presented both a video introduction and slides to every team.
At the end of the introductory phase, the participants were asked to analyze the runtime of a binary search tree implementation with two given inputs to familiarize them with the tool.

\begin{figure}[b!] \centering
\includegraphics[width=0.8\columnwidth, frame]{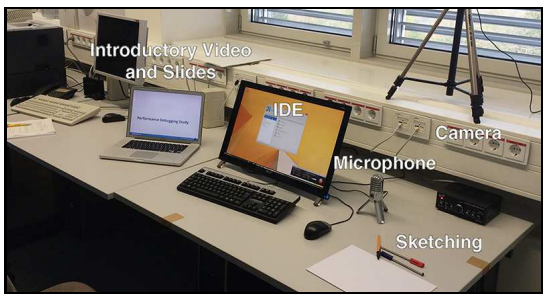}
\caption{Study setup}
\label{fig:setup}
\end{figure} 

\noindent\textbf{2) Locating Bugs:} The teams were prompted to locate and fix four real-life performance bugs having different levels of difficulty. 
After each bug fix, we conducted a structured interview. The laboratory setup is depicted in Figure~\ref{fig:setup}.
The questions for the structured interviews as well as the introductory slides, the video, and the task descriptions for each bug were presented on a laptop next to the computer running the IDE.
These materials are part of the supplementary material~\cite{SupplementaryMaterial}.
Applying pair programming~\cite{Williams00}, one developer was in control of the keyboard and mouse (\emph{driver}), while the other team member acted as a consultant and supervisor (\emph{navigator}).
We switched \textit{driver} and \textit{navigator} after each bug, so that each developer was \emph{driver} for fixing two bugs. 
Participants were asked to verbalize their thoughts, following the \textit{thinking aloud method}~\cite{Boren00}.
To assess how our participants use the profiling tool and navigate through the code, we logged certain user actions, audio-recorded the sessions, and captured the screen.
This enabled us to analyze the understanding process and the communication between the participants.
The pair programming setting in conjunction with the thinking aloud method makes talking more natural compared to a study where the participant is alone.
This setting is based on the \textit{constructive interaction} method proposed by \citeauthor{Miyake86}~\cite{Miyake86}. 

Developers were given sheets of paper and pens that they were allowed to use at any point during the study.
To get insights into the understanding and communication process, they were asked to sketch out the cause of the performance problem and their proposed solution.
This is a realistic scenario because sketches and diagrams play an important role in the daily work of software developers~\cite{Baltes14, Cherubini07}.
We recorded the whole sketching process using an HD video recorder pointed at the area on the table where pen and paper were placed.
This allowed us to analyze developers' understanding of the performance problem using the externalized mental model, the sketch.

\noindent\textbf{3) Questionnaire} In the third and last phase, participants were asked to fill out a questionnaire with questions about demographic data, work experience, experience in different areas (see Section~\ref{sec:participants}), and the usability of the tool.

\subsection{Performance Bugs}

The four performance bugs we used in the second phase were taken from the Apache Commons Collections project (Bugs 1--3) and the \mycode{com.google.common.collect} package of the Guava Libraries project (Bug 4).
These projects were selected because they were mentioned containing interesting performance bugs~\cite{Jin12, Nistor13}, are understandable without specific prior knowledge, and are implemented in Java.
All chosen performance bugs were well-documented and code illustrating the bug was provided. 
Since many performance bugs manifest only for particular inputs~\cite{Nistor13, Jin12}, we prepared test classes to reproduce each bug.
At the beginning of a session, participants were asked to open the test class for the corresponding bug and start the profiling.
They were free to modify or extend the test class if needed.

Based on manual inspection, we classified the bugs as \emph{easy} or \emph{difficult}. In case teams got stuck with one of the bugs, we also prepared advice on how to proceed, which the instructor provided on demand (available as supplementary material~\cite{SupplementaryMaterial}, together with links to the bug descriptions).
 
\noindent\textbf{Bug 1} (easy): 
In class \mycode{ListUtils}, the static method \mycode{retainAll(Collection, Collection)} iterates through 
the first collection and checks whether an element is also contained in the second one; it returns the intersection of the two collections. 
The runtime strongly depends on the dynamic type of the parameters: calling this method with two linear lists results in quadratic runtime complexity, but linear complexity in case the second parameter allows containment checks in constant time (e.g., a hash set). 

\noindent\textbf{Bug 2} (easy):
Class \mycode{SetUniqueList} implements a linear list using a set to maintain uniqueness of elements internally. The performance problem is that the method \mycode{retainAll(...)} calls its superclass implementation as well as \mycode{retainAll(...)} on the internally used set. This leads to two retainment computations, while only one and a few simple maintenance operations would be necessary.

\noindent\textbf{Bug 3} (difficult): 
Within class \mycode{MultiValueMap\$Values}, the omission to override method \mycode{containsAll(...)} causes the superclass method to be called. This method iterates through the collection passed as parameter and checks for every element if it is contained in the current instance of \mycode{Values}. The method \mycode{contains(...)} uses an iterator to loop over all values. This iterator is actually a concatenation of iterators because a \mycode{MultiValueMap} can store multiple values per key. To create such a concatenation of iterators, it takes a complete iteration over all stored value lists. Thus, a call of \mycode{contains(...)} iterates through all values twice and the frequent creation of the iterator is unnecessary.
    
\noindent\textbf{Bug 4} (difficult):  
Class \mycode{RegularImmutableSet} implements a hash table that uses the open addressing approach with linear probing~\cite{Cormen09} to resolve hash collisions. Method \mycode{contains(Object)} consumes much runtime for some special input, which creates long sequences of filled slots so that a lookup operation degrades closer to linear complexity than to constant time (primary clustering issue~\cite{Cormen09}). As a problem of the underlying algorithm, not its implementation, this is not a performance bug according to the definition of \citeauthor{Jin12}~\cite{Jin12}. However, the symptoms the developers observe (i.e., a high runtime consumption) are the same as for a performance bug. We included this example to test how developers react if there is no easy fix available for the performance issue.

\section{Results}
\label{sec:results}

The study sessions lasted between 94 and 164 minutes (mean value: 121 minutes), resulting in over 12 hours of audio and video recordings.
To answer the research questions formulated in Section~\ref{sec:research_questions}, we first analyzed the transcribed answers of the interviews following each bug locating task (Bug 1--4).
To this end, we conducted a \textit{cross-case analysis}~\cite{Seaman99}:
We started with the interview transcripts of the first two teams and wrote down short summaries of participants' statements related to the research questions. 
Then, we compared these statements to determine similarities and differences.
The result of this step was a list with preliminary propositions, based only on the answers of the first two teams.
For each proposition, we wrote down the statements supporting or refuting it.
For all remaining groups, the following process was repeated:
We analyzed the answers of the next team and compiled a list of statements.
Then, we determined if any of those statements supported or refuted the existing propositions.
Supporting statements were added to the corresponding proposition.
In case a statements refuted a proposition, either the proposition was revised or the statement was added to the list of refuting statements for that proposition.
Any additional propositions suggested by the team's statements were added.

The result of this process was a final list of propositions each with a set of supporting and refuting statements.
The transcription of the interviews and a first iteration of the cross-case analysis was conducted by one author.
Then, based on the extracted statements, the resulting propositions were discussed with two other authors.
We decided to drop all propositions with less than four supporting statements. 

For a detailed analysis of code navigation (based on interaction logs) and verbal communication (based on audio and video recordings), we focused on Bug 3.
We assumed that, when starting with the third bug, the participants were familiar with the tool and had some hands-on experience with the Apache Commons Collections library.
Moreover, according to our observations, this bug was the most difficult one to locate and fix. 

\subsection{RQ1 (Navigating and Understanding)}

The first research question (RQ1) covers the navigation process and code comprehension problems related to it. The first and second subquestion (RQ1.1, RQ1.2) were answered based on the results of the cross-case analysis reported in Table~\ref{tab:navigation-cca-1}, while the third one (RQ1.3) was addressed by investigating the recorded interaction logs.

\begin{table}[t]
	\caption{Propositions based on cross-case analysis of interview answers related to RQ1.1 (top) and RQ1.2 (bottom).}
	\label{tab:navigation-cca-1}
	\centering
	\begin{tabularx}{1\columnwidth}{cXc}
		\toprule
		No. & Proposition	& Teams\\		
		
		\midrule
		
		1.1 & The dynamic instance of a method call and connected runtime information are important for navigation. & T1, T3, T4, T5 \\ 
		1.2 & Following high quantities of runtime in the dynamic method call graph is helpful as a navigation strategy. & T1, T2, T3, T6 \\ 
		1.3 & The more complex the performance bug is, the less helpful the provided tool support and information becomes. & T1, T3, T5, T6  \\ 
		
		\midrule 
		2.1 & The integration into the code view provides additional context for the profiling visualization. & T1, T2, T4, T6 \\ 

		2.2 & The overview (list view) was not needed in this setting. & T1, T4, T5  \\ 

		2.3 & The overview (list view) could be used as a starting point for further analyses. & T1, T2, T4 \\ 
		\bottomrule	
					
	\end{tabularx}
\end{table}

\noindent\textbf{RQ1.1} \emph{How was information from the profiling tool or other parts of the IDE used to locate the performance bug?} 

Participants highlighted that locating the bug was particularly supported by using dynamic instances of method calls as links and by the connected runtime information (Prop. 1.1). In this context, the method call visualization that provides runtime information for a specific call was also helpful (T1, T3, T4). Participant P8, for instance, found it useful that ``one could jump to the actual implementation that was executed'' instead of the position where a method has been declared (e.g., in an interface).
This allows developers to follow paths through the code, chasing the largest quantities of runtime consumption as a main navigation strategy (Prop. 1.2).
However, participants also noted that the more complex the bug was, the less helpful the tool support and provided information became (Prop. 1.3). For instance, it was difficult not to get lost in the dynamic call graph (T6).

\emph{Discussion:} These summarized statements confirm that, besides the actual performance information, the performed calls need to be explored to gain an understanding of the performance bug. Although we already provided sophisticated support for navigating along those dynamic calls, the call structure could still become too complex. How to provide better context to prevent developers from getting lost in the call graph remains an open research question.

\noindent \textbf{RQ1.2} \emph{Is the in-situ visualization of the profiling data 
beneficial compared to a traditional list representation?}

In general, the integration of profiling information into the source code was received very positively. In contrast to the list representation, the code provides context necessary to understand the performance information (Prop. 2.1).
Three teams (T1, T4, T5) agreed that the list view of our profiling tool was not required for navigation in the setting of the user study (Prop. 2.2). A reason could be that we gave the performance bugs to our participants ``on a silver platter'' (P4).
Participants remarks indicated that in case they were looking for new performance bugs, the list view would be a good starting point to get an overview on the runtime of all captured methods and classes (Prop. 2.3). 

\emph{Discussion:} In general, integrating source code and performance information seems to be a promising direction to follow. Since our focus was not on detecting new performance bugs, but on locating and understanding known ones, our scenario is, however, limited.
Although the list representation was rated as not necessary for our setting, we cannot conclude that it is superfluous. It rather seems likely that both representations complement each other for different usage scenarios.

\noindent\textbf{RQ1.3} \emph{What navigation strategies do developers pursue to locate a specific performance bug?}

\begin{table}[tbp]
	\caption{Frequency of navigation events discerned by types of navigation.}
	\label{tab:navigation-statistics}
	\centering
	\begin{tabularx}{1\columnwidth}{Xrrrr}
		\toprule
		Type of navigation & \multicolumn{4}{c}{Usage} \\
		& \multicolumn{2}{c}{Total} & \multicolumn{2}{c}{Bug 3}\\	
		\midrule
		Method call visualization & 236 & (25\%) & 98 & (24\%) \\
		Method visualization & 39 & (4\%) & 30 & (7\%) \\
		Class visualization & 7 & (1\%) & 7 & (2\%)\\
		Overview list & 20 & (2\%) & 0 & (0\%)\\
		IDE & 657 & (69\%) & 269 & (67\%)\\
		\bottomrule			
	\end{tabularx}
\end{table}

Every navigation event changing the source code view was recorded. We discern IDE events from different events in the profiling tool. As Table~\ref{tab:navigation-statistics} summarizes for all sessions (Bug 1--4), about two thirds of the navigation events were conducted through standard IDE navigation features.
The other third is related to the profiling tool, dominated by the navigation through method call visualizations. Among the other profiling navigation options, only navigation through tooltips of visualizations next to method declarations played a visible role, while class list visualizations and the list view were used only very rarely.
The distribution of events for Bug 3, which we use in closer detail in the following, is very similar to the overall distribution.
Comparing Bug 3 to the other bugs, a different navigation behavior cannot be observed from these aggregated numbers.

\begin{figure}
\begin{minipage}[b]{1\columnwidth}
\centering
\includegraphics[width=0.88\columnwidth]{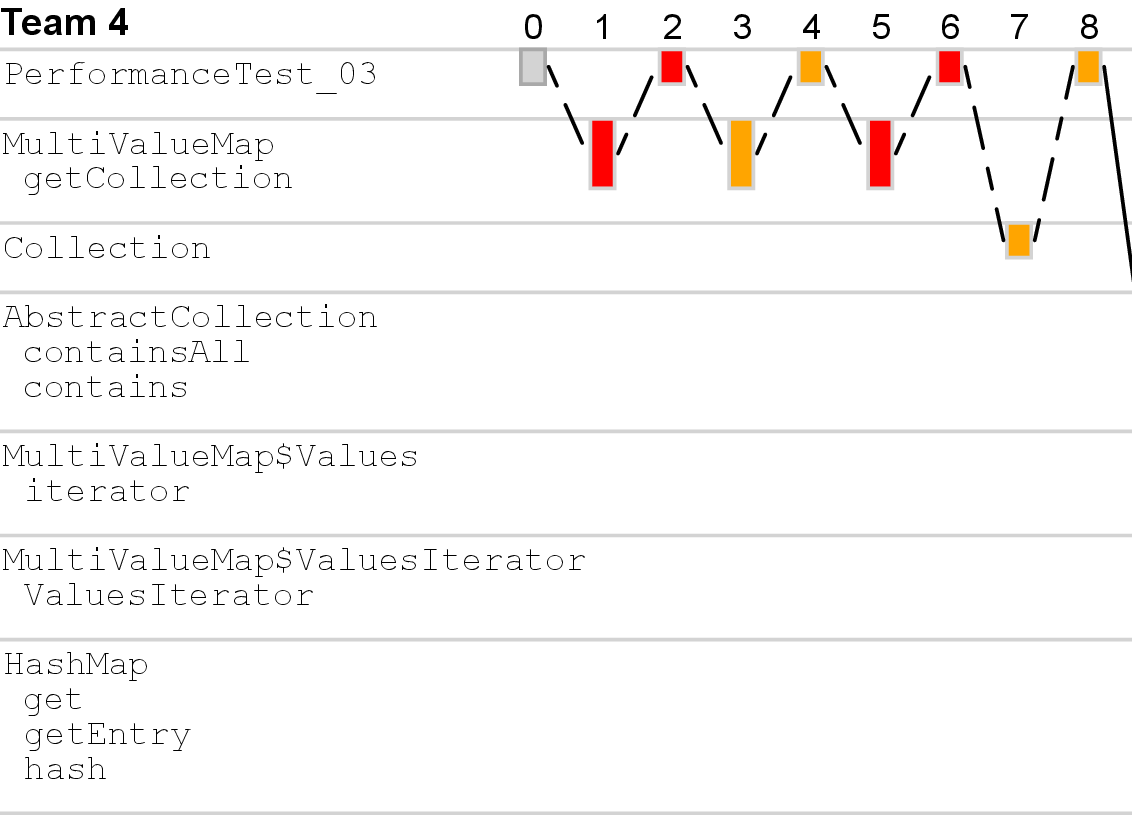}
\subcaption{Strategy 1 (\emph{Toggle})}
\label{fig:strategy1}
\end{minipage}%
\hfil
\begin{minipage}[b]{1\columnwidth}
\centering
\includegraphics[width=0.88\columnwidth]{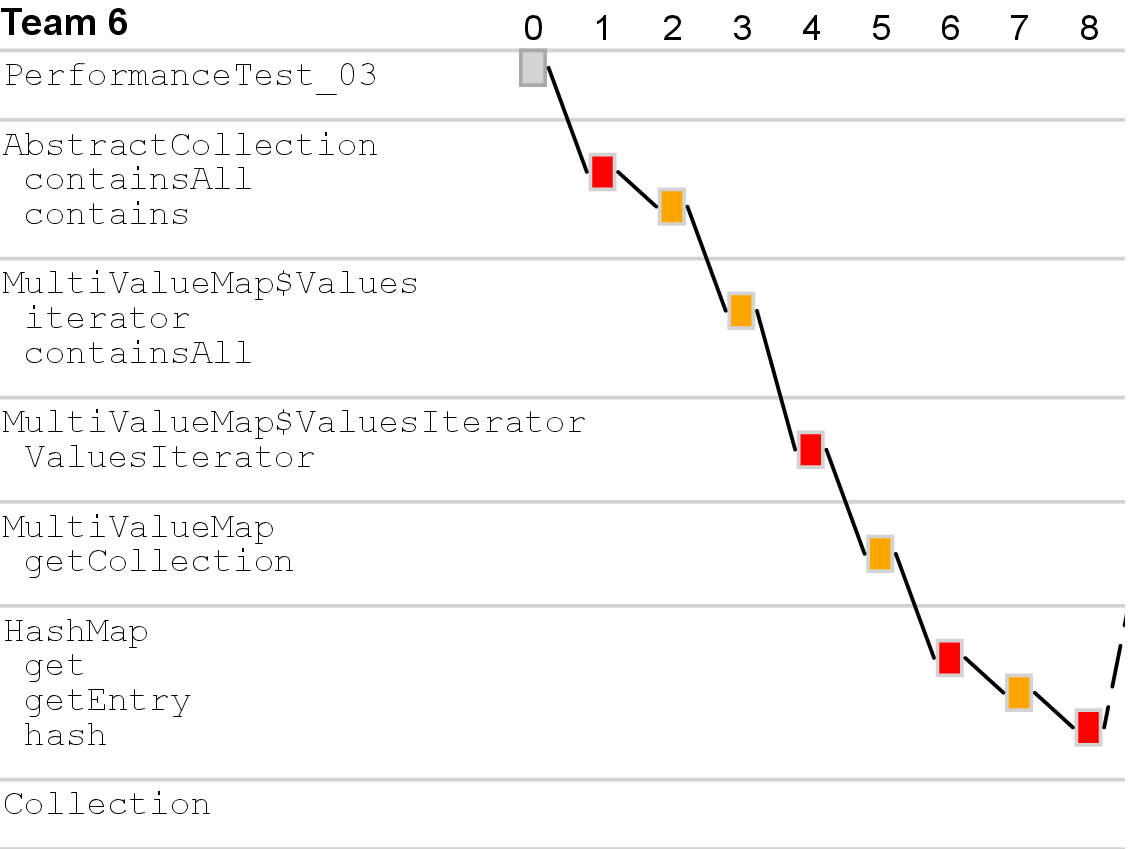}
\subcaption{Strategy 2 (\emph{Path Following})}
\label{fig:strategy2}
\end{minipage}%
\caption{Visualized first phase of the navigation log for T4 and T6 (Bug 3).}
\label{fig:strategies}
\end{figure}

We developed a visualization for the event data to analyze navigation strategies and usage patterns (Fig.~\ref{fig:strategies}). The visualization shows a sequence of navigation events on a discrete time axis. On the left, all visited methods and corresponding classes are listed vertically in order of first visit. The color of each event depends on the time the participants spend in the respective method (yellow: $<$6s; orange: 6--26s; red: $>$26s; gray: unknown).
We chose these split points such that events are distributed uniformly. Links between events indicate the type of navigation used: a solid black line represents navigation through a method call visualization, a solid blue line navigation through a method visualization, and a dashed black line all other types of navigation (IDE navigation and class visualizations). Fig.~\ref{fig:strategies} shows two examples, the complete set of visualizations is provided as part of the supplementary material~\cite{SupplementaryMaterial}.

\begin{table*}[t]
\renewcommand\arraystretch{0.8}
	\caption{Interactions while locating performance bug 3\\(D: During, A: After locating bug, $^{*}$: navigator took over role of driver, Codes: see Table~\ref{tab:codes})} 
	\label{tab:communication}
	\centering
	\begin{tabular}{ccccccccccccccc}
		\toprule
		Team & Time & Success & Driver & Navigator & \multicolumn{8}{c}{Codes} & First Strategy & Sketch \\
		& (min.) & & & &Total & DC+HC & DR+HR & QC+QR & PN+PI & CO & RD+RC+RE & Other & & \\
		\midrule
		\multirow{3}{*}{T1}	 & \multirow{3}{*}{30} & \multirow{3}{*}{\checkmark} &  &  & 165 & 46 & 11 & 28 & 5 & 10 & 11 & 54 & \multirow{3}{*}{1} & \multirow{3}{*}{D} \\
		& &	&  P2 & 	& 45\%& 57\% & 55\% &21\% & 0\% & 20\% & 55\% & 54\% & & \\
		& &	& & P1 & 55\%& 43\% & 45\% & 79\% & 100\% & 80\% & 45\% & 46\% & & \\
		\midrule
		\multirow{3}{*}{T2}	 & \multirow{3}{*}{30} & \multirow{3}{*}{\checkmark} &  &  & 112 & 21 & 19 & 24 & 9 & 6 & 9 & 24 & \multirow{3}{*}{1} & \multirow{3}{*}{A}  \\
		& &	&  P4 & 	& 57\% & 67\% &58\% & 54\% & 11\% &33\% & 56\% & 75\% & & \\
		& &	& & P3 & 43\% & 33\% &42\% & 46\% & 89\% &67\% & 44\% & 25\%	& & \\
		\midrule
		\multirow{3}{*}{T3}	 & \multirow{3}{*}{24} & \multirow{3}{*}{\checkmark} &  &  & 78 & 18 & 13 & 10 & 6 & 7 & 3 & 21 & \multirow{3}{*}{2} & \multirow{3}{*}{A} \\
		& &	&  P5 & 	& 63\% & 83\% & 85\% & 90\% & 0\% & 0\% & 100\% & 52\% & & \\
		& &	& & P6 & 37\% & 17\%  & 15\% & 10\% & 100\%  & 100\%  & 0\%	& 48\%	& & \\
		\midrule
		\multirow{3}{*}{T4}	 & \multirow{3}{*}{35} & \multirow{3}{*}{\checkmark} &  &  & 136 & 24 & 22 & 20 & 15 & 7 & 10 & 38 & \multirow{3}{*}{1} & \multirow{3}{*}{D} \\
		& &	&  P7 & 	& 46\% & 58\% & 68\% & 20\% & 0\% & 29\% & 20\% & 68\% & & \\
		& &	& & P8 & 54\% & 42\%  & 32\% & 80\% & 100\%  & 71\%  & 80\% & 32\% & & \\
		\midrule
		\multirow{3}{*}{T5}	 & \multirow{3}{*}{20} & \multirow{3}{*}{$\circ$} &  &  & 48 & 14 & 9 & 10 & 2 & 0 & 2 & 11 & \multirow{3}{*}{-} & \multirow{3}{*}{D} \\
		& &	&  P10$^{*}$ & & 35\% & 21\% & 44\% & 30\% & 0\% & 0\% & 100\% & 45\% & & \\
		& &	& & P9$^{*}$ & 65\% & 79\%  & 56\% & 70\% & 100\%  & 0\%  & 0\% & 55\% & & \\
		\midrule
		\multirow{3}{*}{T6}	 & \multirow{3}{*}{24} & \multirow{3}{*}{$\times$} &  &  & 40 & 15 & 13 & 1 & 2 & 3 & 0 & 6 & \multirow{3}{*}{2} & \multirow{3}{*}{D}  \\
		& &	&  P12 & & 63\% & 73\% & 77\% & 0\% & 0\% & 0\% & 0\% & 67\% & & \\
		& &	& & P11 & 38\% & 27\%  & 23\% & 100\% & 100\%  & 100\%  & 0\% & 33\% & & \\
		\bottomrule
	\end{tabular}
\end{table*}

The first phase of locating the bug is of particular interest because participants try to understand the circumstances of the performance bug and explore the corresponding code.
A general observation is that the method and method call navigations become rare when the team starts solving the bug. Analyzing the navigation phase at the beginning, we identified two general strategies.
Parts of the teams that started with Strategy 1 later followed Strategy 2 and vice versa (Strategy 1 $\rightarrow$ 2: T4, T5; Strategy 2 $\rightarrow$ 1: T6), some alternating several times (T4, T5). 

\begin{itemize}[leftmargin=*, itemsep=1ex]
\item \textbf{Strategy 1 (\emph{Toggle})}: Some teams frequently switched back and forth between the performance test class and important classes regarding the bug (T1, T2, T4, T5; Fig.~\ref{fig:strategy1}). To this end, they use only IDE functionality. In a way, this strategy is comparable to a breadth-first search while exploring the circumstances of the bug. 
\item \textbf{Strategy 2 (\emph{Path Following})}: Another approach was to follow dynamic calls with high runtime consumption through the method call visualizations until reaching a method with high self time (T3, T6; Fig.~\ref{fig:strategy2}).
This is comparable to following a path on the dynamic call graph to its end, similar to a depth-first search. 
\end{itemize}

\emph{Discussion:} Given the heterogeneous characteristics of participants (see Table~\ref{tab:participants}), we were surprised to find clear, repeated strategies. Since most of the interaction subsequences can be assigned unambiguously to either of the strategies, these two strategies seem to be applicable rather universally. None of the strategies is obviously dominating the other, neither in regard of frequency nor order. Further, we were not able to detect other strategies on a comparably generic level.

\begin{normalbox}
\noindent \textbf{RQ1} (Navigating and Understanding): \emph{How do developers navigate and what information and representation is supportive for locating a performance bug?}

\noindent\emph{Summary}: Dynamic instances of method calls and related runtime consumption are central sources of information for locating a performance bug. Visually integrating this information into the code can be beneficial and, in our scenario, replaced a list representation of the performance information. We identified two main navigation strategies that occured in alternating patterns: (i) to toggle between performance test and production code, (ii) to follow paths of dynamic calls.
\end{normalbox}

\begin{table}[tbp]
	\caption{Codes used in Table~\ref{tab:communication}}
	\label{tab:codes}
	\centering
	\begin{tabularx}{\columnwidth}{cX}
		\toprule
		Code	& Description \\
		\midrule
		DC		& Describes source code (e.g., data structure, architecture, algorithm)\\
		HC		& Expresses hypothesis about how the source code works.\\
		DR		& Talks about runtime or refers to profiling data  \\
		HR 		& Expresses hypothesis about runtime \\
		QC		& Question regarding source code (e.g., data structure, architecture, algorithm)\\
		QR		& Questions that explicitly mentions the runtime or profiling data \\
		PN		& Prompt to navigate (e.g., ``go to this method'') \\
		PI		& Prompt to implement (e.g., ``you have to write \texttt{for (int i: ...)}'') \\
		CO		& Disrupting comment (e.g., ``Stop! We have to...'') \\
		RD		& Reads documentation/source code comment aloud \\
		RC		& Read source code aloud \\
		RE 		& Reference to source code (``There is the problem.'') \\
		\bottomrule	
	\end{tabularx}
\end{table}

\subsection{RQ2 (Understanding and Communicating)}
\label{sec:communicating}

The second research question (RQ2) focuses on the understanding process while locating a bug, expressed by the communication that takes place.
To answer the first subquestion (RQ2.1), we transcribed the audio recordings from all sessions for Bug 3.
Then, one of the authors employed an open coding approach~\cite{Corbin08, Charmaz14} and iteratively assigned codes to each statement of the participants:
In a first initial coding phase, the transcripts were coded on a statement level and memos were utilized to structure emerging patterns and similar codes.
In a second, focused, coding phase the codes were revised or merged where applicable, concentrating on how the team members interacted with each other.
During this phase, we iterated over the transcripts several times.
For the second subquestion (RQ2.2), we again utilized the transcribed recordings, but we also integrated results from the cross-case analysis.

\noindent\textbf{RQ2.1} \emph{How do developers communicate with each other when locating a performance bug?}

Table~\ref{tab:communication} shows the results for the twelve codes we base our analysis on; Table~\ref{tab:codes} provides descriptions for the codes.
A complete list with all codes and the data for each session are available as supplementary material~\cite{SupplementaryMaterial}.
The locating phase for Bug 3 lasted between 20 and 35 minutes.
Four teams created a working fix, but all teams needed help by the instructor.
Team T5 created a fix that altered the semantics of the data structure and, for team T6, the instructor had to reveal the complete solution before they were able to implement a fix.
In the following, we will describe the communication behavior of the teams based on the transcripts and codes. 

Most teams expressed their first hypothesis about a possible performance problem in the first half of the session and two teams (T1, T4) in the second half.
The role of the navigator differed across the teams:
In half of the teams (T1, T4, T5), the navigator was very active, asking questions about the source code or profiling data (value in column QC+QR $>$75\%), prompting the driver to navigate to certain methods (PN) or dictating when the driver was writing source code (PI).
In T2, the communication was rather balanced between driver and navigator.
The passive navigators in T3 and T6 mostly observed and infrequently interacted with the driver.  
In all except one team (T5), the navigator expressed interrupting comments (CO), e.g., to indicate his notion of a piece of code.
In most cases, the navigator responded to these comments.
However, two teams stand out: In T3, the driver almost completely ignored the navigator's comments and continued with his source code descriptions and hypotheses (DC+HC 83\%, DR+HR 85\%).
We observed similar behavior for T5, where the navigator took over the role of the driver and finally implemented the solution himself.
One reason for the dominant driver in T3 and the rather dominant navigator in T5 could be a difference in expertise:
Both the driver of T3 and the navigator of T5 rated themselves as more experienced in OOP, collections, and fixing performance bugs compared to their teammates (see Table~\ref{tab:participants}).
When understanding the program, it was quite common to point to source code and read it or the related documentation aloud (RE+RC+RD).
The role of the navigator is often described as thinking strategically instead of focusing on the implementation~\cite{Chong07}.
However, in our study, driver and navigator mostly worked on the same level of abstraction, for instance when reading source code aloud or talking about it (DC+HC).

\emph{Discussion:} Unlike for the navigation strategies, we could not detect as clear communication strategies.
Probably depending on the different level of expertise of the participants and the composition of the teams, the communication patterns diverge more between the teams.
Most teams formulated a hypothesis for the cause of the performance bug relatively early, i.e., in the first half of the session.
They discussed the architecture and algorithms related to the performance bug while working together on their hypothesis.
However, in two teams a dominant team member lead the interaction.

\begin{table}[tbp]
	\caption{Propositions based on cross-case analysis of interview answers related to RQ2.2.}
	\label{tab:navigation-cca-2}
	\centering
	\begin{tabularx}{1\columnwidth}{cXc}
		\toprule
		No. & Proposition	& Teams\\		
		
		\midrule

		3.1 & Sketches are a useful tool for explaining a performance bug, but context information is needed to understand them afterwards. & T1, T2, T4, T6 \\ 
				
		3.2 & Sketches are a suitable documentation means (if ``polished'' enough). & T1, T2, T3, T4, T6 \\ 

		3.3 & If and how much sketching occurs depends on the sketching experience of the developers. & T1, T2, T5, T6 \\ 

		3.4 & A common sketch vocabulary is needed in the team. & T1, T2, T5 \\ 

		3.5 & More complex problems or data structures are more likely to be sketched. & T1, T2, T4, T6 \\ 
						
		3.6 & Sketches can be used to explain dynamic aspects of a program. & T1, T2, T4\\		
		\bottomrule			
	\end{tabularx}
\end{table}

\noindent\textbf{RQ2.2} \emph{Could sketches help understand and communicate a performance bug?}

\begin{figure}[tbp] \centering
	\includegraphics[width=0.55\columnwidth, valign=c]{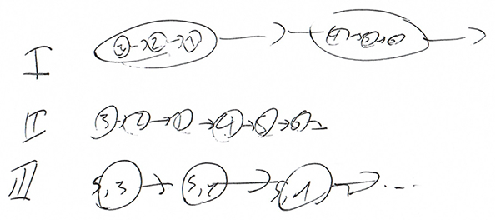}
	\hspace{1.5em}
	\includegraphics[width=0.3\columnwidth, valign=c]{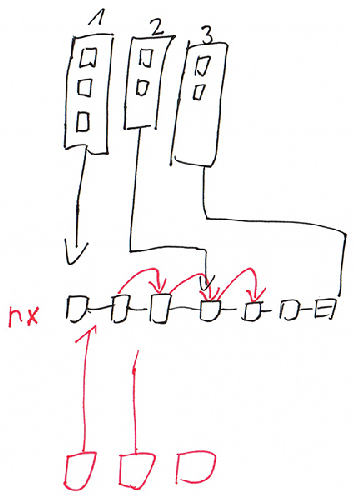}
	\caption{Sketches to explore alternatives (left) and dynamic behavior (right)}
	\label{fig:sketch}
\end{figure}

Four teams created a sketch spontaneously while locating Bug 3, the other two were asked to sketch the problem and their solution afterwards (see Table~\ref{tab:communication}).
Some teams sketched only the static structure of a \mycode{MultiValueMap} (T1, T4, T5), others expressed also dynamic aspects like the execution of method \mycode{contains(...)} (T2, T3, T6; see Fig.~\ref{fig:sketch}).
Expressing dynamic behavior was also mentioned several times during the interviews (see Table~\ref{tab:navigation-cca-2}, Prop. 3.6).
Despite only depicting the static structure, teams T1 and T4 referenced their sketch several times during the session; team T4 used it to explore alternative hypotheses about the data structure (Fig.~\ref{fig:sketch}).
In all teams that created sketches during the location phase, the navigator sketched the data structure and explained aspects of it to the driver.

During the interviews, developers rated sketches as being a useful tool for explaining performance bugs to someone else, especially when trying to understand the data structures in use (e.g., the \mycode{MultiValueMap}).
However, it is not only the artifact that matters but the whole context of creating a sketch (e.g., order of creation, related source code, conversation) (Prop. 3.1).
This makes it difficult to use the sketches afterwards for explaining the problem to someone else, but if they are edited and ``polished'' enough, they may be used for documenting a bug fix (Prop. 3.2).
Generally, the participants noted that the sketching practice depends on the experience of the developers (Prop. 3.3).
One participant even reported a ``training effect'' while sketching the team's solution for Bug 2.   
Furthermore, participants pointed at the need of a common sketch vocabulary for the team (Prop. 3.4).
This vocabulary can be informal and may emerge during a meeting.
P3, for instance, noted that in his team, they agreed on the convention that ``the circle with the dots is always a \mycode{HashSet}''.
Also, more complex problems or data structures are more likely to be sketched (Prop. 3.5).
Several participants reported that there is a point when it becomes too difficult to keep a problem or a data structure in the mind.
This is when a sketch is created and hence the developer's mental model gets externalized.

\emph{Discussion:} Although imposed by the study procedure, sketching was considered mostly positive as an aid for explaining a performance bug. The natural use of sketching performed by the navigator in four teams during the location process suggests that it is an appropriate tool for pair programming scenarios. It becomes less clear, however, whether sketching would also be a useful externalization for a single developer when trying to solve a complex performance bug. 

\begin{normalbox}
\noindent \textbf{RQ2} (Understanding and Communicating): \emph{How do developers try to understand and explain the causes of performance bugs?}

\noindent\emph{Summary}: Most pair programming teams formulated a hypothesis for the cause of the performance bug early and discussed the architecture and algorithms related to the bug.
Within a team, the role of the navigator can range from active (posing questions, commenting) to passive (mostly observing).
Sketches may be a useful tool to explain performance problems to co-workers.
\end{normalbox}

\section{Threats to Validity}
\label{sec:validity}

By inviting participants with diverse backgrounds (research assistants and graduate students with and without industry experience as well as professional software developers) and choosing four real-world performance bugs from widely used open source projects, we tried to limit threats to the external validity of our study.
However, the focus on performance bugs in collection libraries and the limited size of our subject sample limit the generalizability of our findings. 
To counter participants' low experience with profiling tools, we conducted a detailed introduction into our sampling approach and let them experiment with our tool before giving them the first performance bug.
The fact that some participants had low experience with the IntelliJ IDEA IDE may also affect the validity of our study.
We tried to mitigate this threat by concentrating on the third bug where we assumed that the developers had time enough to gather experience with the IDE.
The same applies for the collection libraries in which our participants had to fix the performance bugs.

Since our study sessions took quite long, the internal validity may be affected by fatigue effects, especially for the last bug.
Another threat to internal validity may be the support by one of the authors during the study.
To mitigate this threat, we only gave the participants hints when they got stuck.
Furthermore, the hints had been prepared before conducting the study and were given in the same order to every team.
The fact that most teams did not work together before our study could also affect the results.
Thus, we concentrated on the third bug for our analysis, when the team members had at least some time to get used to each other and the environment.

The reliability of our results may be limited due to the fact that only one author conducted each the initial cross-case analysis and the coding of the interview transcripts.
We tried to mitigate this ``lone researcher bias''~\cite{Burnard08} by later discussing the results among the authors.

\section{Related Studies}
\label{sec:related_work}

As mentioned earlier, we are not aware of any previous study investigating in particular how developers locate performance bugs.
However, studies have been performed that generally address change tasks, program comprehension, feature location, and code navigation.
Beyond that, studies exist that lead to first tools to automatically detect performance bugs in software projects.
We summarize related work and link the results to ours.

\citeauthor{Jin12}~\cite{Jin12} conducted a study of 109 real-world performance bugs to extract certain characteristics.
Based on their findings, they implemented a rule-based performance bug detection and found many previously unknown performance problems in several open source projects.
\citeauthor{Nistor13}~\cite{Nistor13} used the bugs that \citeauthor{Jin12} collected to identify how these bugs depend on loops and implemented an automated oracle for performance bugs.
Using this oracle, they found 42 new performance bugs in 9 open source Java projects. We used three of those bugs in our study.

\citeauthor{Roehm12}~\cite{Roehm12} conducted an observational study with 28 professional software developers.
They found that developers prefer face-to-face communication over documentation, which supports our choice to conduct the study in a pair programming setting.
Furthermore, the bug locating phase of our study models the problem-solution-test work pattern they described: for bug fixing tasks, developers first identify the problem, then search for and apply a solution, and finally test the correctness of the solution.
Similar to \citeauthor{Sillito08}~\cite{Sillito08} and \citeauthor{Ko07}~\cite{Ko07}, \citeauthor{Roehm12} found that while developers comprehend software, they ask and answer questions and test hypotheses about application behavior.
We can partially confirm this for our setting (see Section~\ref{sec:communicating}).
Insufficient investigation of the code prior to a change task may lead to an ``ignorant surgery''~\cite{Parnas94} and thus to less successful solutions~\cite{Robillard04}.
We found that, before implementing a fix, developers extensively navigate and discuss the source code that they think is related to a performance problem.

\citeauthor{Baltes14}~\cite{Baltes14} found in a study involving 394 software practitioners that sketches and diagrams play an important role for understanding and explaining source code.
In accordance to that, \citeauthor{Cherubini07}~\cite{Cherubini07} name understanding and communicating as two of the most important motivations for creating sketches.
In our study, the participants found sketches to be a useful tool for explaining performance bugs.

Most existing studies found no significant influence of developers' personality on pair programming effectiveness~\cite{Salleh09}.
However, developers' expertise can affect pair programming sessions: 
\citeauthor{Chong07} conducted an ethnography study of professional pair programmers from two software development teams.
They found that in teams with different levels of expertise for driver and navigator, the less knowledgeable developer had a tendency to become more passive, letting the expert dominate the interaction.
We also observed this behavior in two of six teams.
Further, they reported that during pair programming, developers usually discuss issues on the same level of abstraction, being in line with \citeauthor{Bryant08}'s findings~\cite{Bryant08}.
We can confirm this with our study.

\citeauthor{Lawrance13}~\cite{Lawrance13} suggest an information foraging perspective on debugging where, instead of testing hypotheses, the developers follow a scent while navigating through the code. In an evaluation with twelve developers, they found that a prediction of developers' code navigation behavior based on this theory was more accurate than predictions based on other models. Also, \citeauthor{Beck15}~\cite{Beck15} describe locating specific code as an iterative foraging process including steps like search, filter, and follow relations. The latter step is similar to exploring the dynamic call graph (Prop. 1.2).
In particular, they investigated how developers locate features related to functional bugs and feature requests in a qualitative user study with 20 developers.
In conformance to our results, for locating a specific feature, they confirm the importance to read the source code and the integration of code with additional information. 

\section{Conclusion}
\label{sec:conclusion}

To better understand how developers locate performance bugs, we conducted a qualitative user study observing twelve developers in a controlled setting. Working in pairs, they tried to solve four different performance issues from two open source Java systems. They used a profiling tool that not only represents profiling results as a list, but also integrates profiling information visually into the source code view of the IDE.
The results of the study show that this integration was well-received by the developers. The in-situ visualization largely replaced the list representation within the investigated scenario, although the list might be helpful as a starting point of further performance analyses.
Hence, adding in-situ visualizations would be a promising opportunity to further improve existing profiling tools.
In general, following dynamic calls is an important means of navigation for locating a performance bug. In particular, we identified two navigation strategies: (i) switching back and forth between performance test and production code, (ii) tracing paths through the dynamic call graph. 
A next step is to investigate to what extent existing tools already support these strategies and derive suggestions for enhancements. 
In our study, we found diverse patterns in the communication between developers working in pairs.
In future work, we plan to further analyze the interaction between the developers while locating and fixing performance bugs.
One interesting aspect could be the knowledge transfer that takes place during these tasks~\cite{Plonka15, Zieris14}.
Sketches could be an appropriate medium to explain performance bugs, in particular, describing complex problems.

\section*{Acknowledgment}

The authors thank all developers participating in the study. Fabian Beck is indebted to the Baden-W\"urttemberg Stiftung for the financial support of this research project within the Postdoctoral Fellowship for Leading Early Career Researchers.

\bibliographystyle{IEEEtranN}
%
\bibliography{literature,additional}

\end{document}